\documentclass[acmtog, nonacm]{acmart}

\usepackage{booktabs} 
\usepackage{gensymb}
\usepackage{siunitx}
\usepackage{etoolbox}
\makeatletter
\patchcmd\@combinedblfloats{\box\@outputbox}{\unvbox\@outputbox}{}{\errmessage{\noexpand patch failed}}

\graphicspath{{"figures/"}}

\acmJournal{TOG}

\DeclareRobustCommand{\thinskip}{\hskip 0.1em\relax}
\def\emdash{---}
\def\d@sh#1#2{\unskip#1\thinskip#2\thinskip\ignorespaces}
\def\Dash{\d@sh\nobreak\emdash}
\def\Ldash{\d@sh\empty{\hbox{\emdash}\nobreak}}
\def\Rdash{\d@sh\nobreak\emdash}

\begin{document}

\title{Gaze-Contingent Ocular Parallax Rendering for Virtual Reality}

\author{Robert Konrad}
\affiliation{%
  \institution{Stanford University}
}
\email{rkkonrad@stanford.edu}
\author{Anastasios Angelopoulos}
\affiliation{%
  \institution{Stanford University}
}
\email{nikolasa@stanford.edu}
\author{Gordon Wetzstein}
\affiliation{%
  \institution{Stanford University}
}
\email{gordon.wetzstein@stanford.edu}

\renewcommand{\shortauthors}{Konrad, R. et al}

\begin{abstract}
Immersive computer graphics systems strive to generate perceptually realistic user experiences. Current-generation virtual reality (VR) displays are successful in accurately rendering many perceptually important effects, including perspective, disparity, motion parallax, and other depth cues. In this paper we introduce ocular parallax rendering, a technology that accurately renders small amounts of gaze-contingent parallax capable of improving depth perception and realism in VR. Ocular parallax describes the small amounts of depth-dependent image shifts on the retina that are created as the eye rotates. The effect occurs because the centers of rotation and projection of the eye are not the same. We study the perceptual implications of ocular parallax rendering by designing and conducting a series of user experiments. Specifically, we estimate perceptual detection and discrimination thresholds for this effect and demonstrate that it is clearly visible in most VR applications. Additionally, we show that ocular parallax rendering provides an effective ordinal depth cue and it improves the impression of realistic depth in VR.
\end{abstract}


\keywords{computational displays, virtual reality, augmented reality, eye tracking, gaze-contingent rendering}

\begin{teaserfigure}
  \centering
  \includegraphics[width=0.9\textwidth]{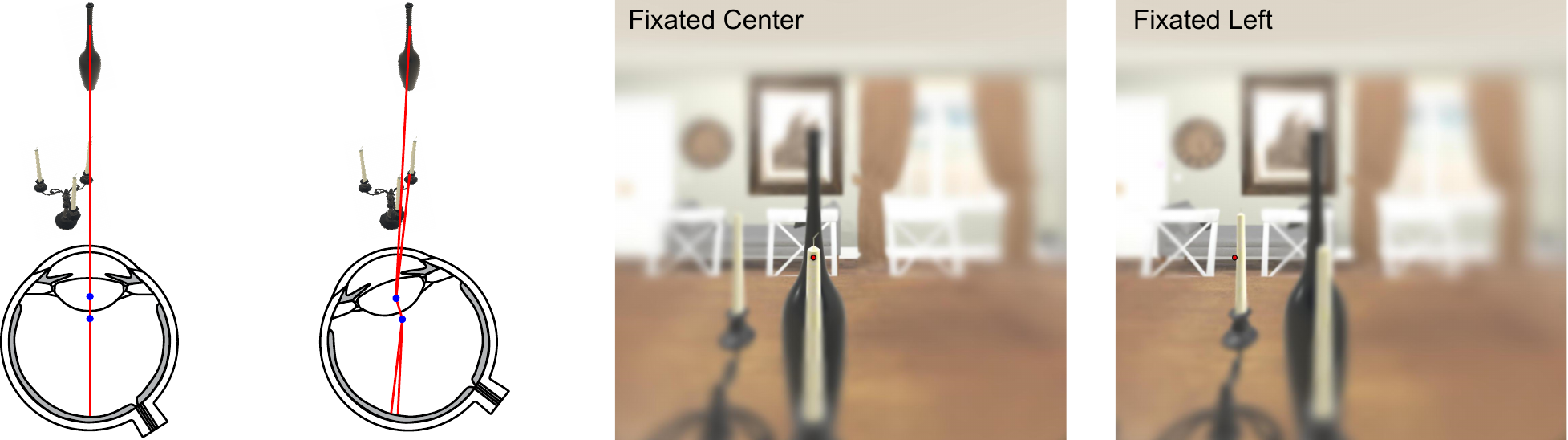}
  \caption{The centers of rotation and projection of the eyes are not the same. As a consequence, small amounts of parallax are created in the retinal image as we fixate on different objects in the scene. The nodal points of the eye, representing the centers of projection, are shown as small blue circles on the left along with a ray diagram illustrating the optical mechanism of ocular parallax. Simulated retinal images that include the falloff of acuity in the periphery of the visual field are shown on the right. As a user fixates on the candle in the center of the scene (center, red circle indicates fixation point), the bottle is partly occluded by the candle. As their gaze moves to the left, ocular parallax reveals the bottle behind the candle in the center (right). Ocular parallax is a gaze-contingent effect exhibiting the strongest effect size in near to mid peripheral vision, where visual acuity is lower than in the fovea. In this paper, we introduce ocular parallax rendering for eye-tracking-enabled virtual reality displays, and study the complex interplay between micro parallax, occlusion, visual acuity, and other perceptual aspects of this technology in simulation and with a series of user experiments.}
	\label{fig:teaser}
\end{teaserfigure}

\maketitle

\section{Introduction}
\label{sec:introduction}

Immersive computer graphics systems, such as virtual reality (VR) displays, aim at synthesizing a perceptually realistic user experience. To achieve this goal, several components are required: interactive, photorealistic rendering; a high-resolution, low-persistence, stereoscopic display; and low-latency head tracking. Modern VR systems provide all of these capabilities and create experiences that support many, but not all, of the monocular and binocular depth cues of the human visual system, including occlusions, shading, binocular disparity, and motion parallax. The support of focus cues (e.g. accommodation and retinal blur) has also received attention in research and industry over the last few years. In this paper, we study a depth cue of human vision that has not been discussed in the context of virtual reality and that may help further improve depth perception and perceptual realism: ocular parallax.


The centers of rotation and projection in the human eye are not the same. Therefore, changes in gaze direction create small amounts of depth-dependent image shifts on our retina \Dash an effect known as ocular parallax. This depth cue was first described by Brewster~\shortcite{Brewster:1845} and it has been demonstrated to produce parallax well within the range of human visual acuity~\cite{Hadani:80,Mapp:1986,Bingham:1993}. Interestingly, species as diverse as the chameleon and the sandlance critically rely on this depth cue to judge distance~\cite{Land:1995,PETTIGREW1999421}.


To render ocular parallax into a VR/AR experience, eye tracking is required. Conveniently, many emerging wearable display systems already have eye tracking integrated, either to support foveated rendering~\cite{Guenter:2012,Patney:2016}, accurate registration of physical and digital images in AR, or other gaze-contingent display modes. With eye tracking available, there is no additional computational cost to integrate ocular parallax. The perspective of the rendered image simply changes depending on the gaze direction. However, the magnitude of depth-dependent motion induced by ocular parallax rendering increases in the periphery of the visual field, where visual acuity is lower than in the fovea. Moreover, the resolution offered by current-generation VR displays is well below the visual acuity of human vision and it is not clear if the subtle ocular parallax effect is perceivable in VR at all. 


To further our understanding of ocular parallax and its perceptual effects in VR, we thoroughly analyze the tradeoffs between perceived parallax, visual acuity, and disparity for near-eye displays. We build a prototype gaze-tracked VR display, conduct a series of user experiments that quantify effect sizes of ocular parallax rendering, and measure its impact on depth perception and the user experience in general. We find that detection thresholds for ocular parallax rendering are almost an order of magnitude lower than the visual acuity at the same extrafoveal locus, which verifies that our sensitivity to small amounts of differential motion is well below the acuity limit, especially in the periphery of the visual field~\cite{Mckee:84}. We also show that the relative ocular parallax of objects with respect to a background target can be discriminated accurately even for relatively small object distances that fall well within the depth ranges of most virtual environments. 
Furthermore, we show that ocular parallax rendering provides an effective ordinal depth cue, helping users better distinguish the relative depth ordering of a scene, but that it does not necessarily benefit absolute, or metrical, distance estimates to objects.
Finally, we show that ocular parallax rendering improves the impression of realistic depth in a 3D scene. For no additional computational cost, ocular parallax rendering has the potential to improve both depth perception and perceptual realism of eye-tracked AR/VR systems.


Specifically, we make the following contributions:
\begin{itemize}
	\item We introduce gaze-contingent ocular parallax rendering for VR display systems.
	\item We design and conduct user experiments to quantify detection and discrimination thresholds of ocular parallax rendering.
	\item We design and conduct user experiments to quantify the effectiveness of ocular parallax rendering as both an ordinal and absolute depth cue.
	\item We conduct a user experiment that demonstrates improved perceptual realism using ocular parallax rendering.
\end{itemize}

%
	%

\section{Related Work}
\label{sec:related}

\paragraph{Depth Cues}

Human depth perception relies on a variety of cues \cite{Palmer:1999,Howard:2002}. Many of these cues are pictorial and can be synthesized using photorealistic rendering techniques, including occlusions, perspective foreshortening, texture and shading gradients, as well as relative and familiar object size. Unlike conventional 2D displays, head-mounted displays (HMDs) use stereoscopic displays and head tracking and can thus support two additional visual depth cues, disparity and motion parallax, as well as one oculomotor cue, vergence. Emerging near-eye displays also support visual focus cues like retinal blur and chromatic aberrations, which in turn drive accommodation, another oculomotor cue (see discussion below). All of these cues are important for human depth perception to varying degrees depending on the fixation distance~\cite{Cutting:1995}. Studying visual cues, such as disparity~\cite{Didyk:2011} or motion parallax~\cite{Kellnhofer:2016}, and their impact on computational display applications has been an integral part of graphics research. In this work, we explore ocular parallax as another visual cue that may improve the user experience in immersive computer graphics applications.

\paragraph{Ocular Parallax}

Ocular parallax describes the change in perspective as the eye rotates, primarily due to the user fixating on different parts of the scene. This visual cue is well known~\cite{Brewster:1845} and has a measurable effect on depth perception~\cite{Mapp:1986,Bingham:1993,Kudo:98,Kudo:99}. Similar to other monocular visual cues, such as retinal blur and chromatic aberration, the change of the retinal image caused by ocular parallax may be small. Nonetheless, supporting all of these cues with an HMD can improve visual comfort~\cite{Hoffman:2008}, perceived realism, and the user experience as a whole.

\citet{Kudo:00} discuss gaze-contingent optical distortions in head-mounted displays (HMDs) and attribute them in part to ocular parallax. This effect is commonly known as ``pupil swim''. However, they did not propose ocular parallax rendering for HMDs or study its perceptual effects with HMDs. Building on emerging head-mounted displays with eye-tracking technology, to our knowledge we are the first to propose ocular parallax as a gaze-contingent rendering mode for VR/AR and evaluate its perceptual implications with a series of user experiments.

\paragraph{Gaze-contingent and Computational Displays}

Eye tracking enables gaze-contingent rendering techniques that adapt effects like magnification, stylization, or geometric level-of-detail to the user's viewpoint~\cite{Duchowski:2004}. Gaze-contingent rendering is becoming an integral part of modern near-eye display systems, enabling techniques such as foveated rendering~\cite{Guenter:2012,Patney:2016}, and gaze-contingent varifocal~\cite{ShengLiu:2008,Konrad:2015,Johnson:16,Padmanaban2183,Dunn:2017} or multifocal~\cite{Rolland:00,Akeley:2004,Mercier:2017} displays. Although rendering accommodation-dependent effects, such as chromatic aberrations~\cite{Cholewiak:2017} and blur at depth edges~\cite{Marshall:96,Zannoli:2014:CBA}, have not been directly evaluated with eye-tracked displays, these techniques could be optimized by tracking the user's gaze or accommodation. Ocular parallax rendering is complimentary to  these techniques and could be integrated, without computational overhead, into conventional HMDs with eye tracking and optionally combined with other gaze-contingent rendering algorithms.

Other than the proposed method, the only techniques described in the literature that inherently provide ocular parallax cues are near-eye multifocal displays~\cite{Love:09,Narain:2015,Hu:14,Llull:15}, light field displays~\cite{Lanman:2013,Hua:14,Huang:2015}, and holographic displays~\cite{Maimone:2017,Padmanaban:2019:olas}. However, the effect of ocular parallax or lack thereof has not been investigated in any of the aforementioned technologies. In fact, ocular parallax in multifocal displays is often undesirable because it reveals misalignments between the virtual image planes; Mercier et al.~\shortcite{Mercier:2017} proposed a multifocal display that effectively removes the ocular parallax cue by shifting the decomposed layers according to the tracked pupil position.

\section{Ocular Parallax}
\label{sec:ocular_parallax}

In this section, we discuss schematic models of the human eye with a specific focus on how they model the centers of projection and rotation. 
Moreover, we discuss and analyze the trade-off between ocular parallax and image sharpness, accounting for the decrease in peripheral visual acuity as well as retinal blur due to accommodation.



\paragraph{\bf Eye Models}

\begin{figure}[t!]
	\centering
		\includegraphics[width=0.6\columnwidth]{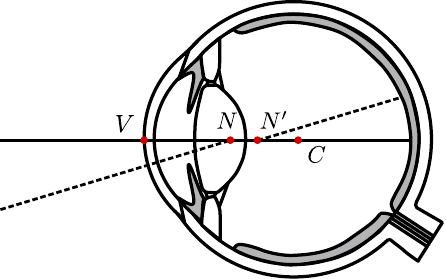}
		\caption{Illustration of a schematic eye, including the front and rear nodal points $N,N'$, the center of rotation $C$, and the anterior vertex of the cornea $V$. The nodal points are two parameters of a thick lens model that refracts light rays as depicted. The exact locations of these points depend on the schematic eye model used.}
		\label{fig:eye}
\end{figure}

Complex optical systems with multiple refractive surfaces can be reduced to six cardinal points that fully define the Gaussian imaging and magnification properties of the system. These cardinal points include the front and rear focal points, the front and rear principle points, and the front and rear nodal points, $N$ and $N'$. For the purpose of modeling ocular parallax, we only require the front nodal point $N$, which is the center of projection of the eye, as well as the center of rotation $C$ (see Figure ~\ref{fig:eye}). Several schematic eye models have been proposed in the literature, each listing slightly different values for the cardinal points~\cite{Atchison:2017}. Some of the most popular models include the Gullstrand number 1, the Gullstrand-Emsley, and the Emsley reduced eyes, which are models of decreasing complexity. We outline the location of their respective nodal points and centers of rotation in Table~\ref{tbl:eyemodels}. The center of rotation of the eye was measured to be 14.7536~mm from the cornea, on average, for emmetropic subjects~\cite{Fry:1962}.

Although the specific locations of the cardinal points are slightly different for each eye model, the distance between center of rotation $C$ and center of projection $N$ is 7--8~mm in all cases. Note that nodal points for the Gullstrand number 1 and Gullstrand-Emsley models are accommodation dependent, i.e. the nodal points move slightly toward the cornea when the eye accommodates to close distances. However, in most current-generation VR/AR systems, the focal plane of the display is fixed. For example, the focal planes of the Oculus Rift and Microsoft Hololens are approximately 1.3~m and 2~m\footnote{\url{https://docs.microsoft.com/en-us/windows/mixed-reality/comfort}} in front of the user, respectively. Since users will accommodate at that fixed focal distance~\cite{Padmanaban2183}, we use the relaxed setting of the popular Gullstrand-Emsley eye model for all experiments in this paper, i.e. $NC=7.6916$~mm.

\begin{table}[t!]
  \centering
  \caption{Overview of parameters of popular schematic eye models: Gullstrand Number 1 (Gull. 1), Gullstrand-Emsley (Gull.-Ems.), and Emsley reduced (Ems.). The distances of front and rear nodal points $N,N'$ are listed in mm with respect to the anterior vertex of the cornea $V$ for the relaxed and accommodated (acc.) state.}
  \renewcommand{\arraystretch}{1.2}
  \begin{tabular}{|l|c|c|c|c|c|}
    \hline
			& Gull. 1 		& Gull. 1 			& Gull.-Ems. 	& Gull.-Ems. 	& Ems.	\\
			& Relaxed 		& Acc.					& Relaxed 		& Acc. 				& 			\\ \hline
	VN	& 7.078				& 6.533					& 7.062				& 6.562 			&	5.556	\\ \hline
	VN'	& 7.331				& 6.847					& 7.363				& 6.909 			&	5.556	\\ \hline
  \end{tabular}
  
	\label{tbl:eyemodels}
\end{table}

\paragraph{\bf Parallax and Acuity}

We can model the amount of perceived ocular parallax expected in various viewing conditions. Similar to the illustration in Figure~\ref{fig:teaser}, we simulate two points that are directly in front of the eye at some relative distance to one another. As the eye rotates, the retinal images of these points will be perceived at an increasing eccentricity, or distance from the fovea, measured in degrees of visual angle. The larger the eccentricity, the larger the parallax, or relative distance, of the points on the retina. However, the density of photoreceptors, in particular the cones, decreases rapidly with increasing eccentricity. Thus, while one would expect a larger amount of parallax in the periphery, the threshold to perceive it there is higher due to the falloff in visual acuity.

\begin{figure}[t!]
	\centering
		\includegraphics[width=\columnwidth]{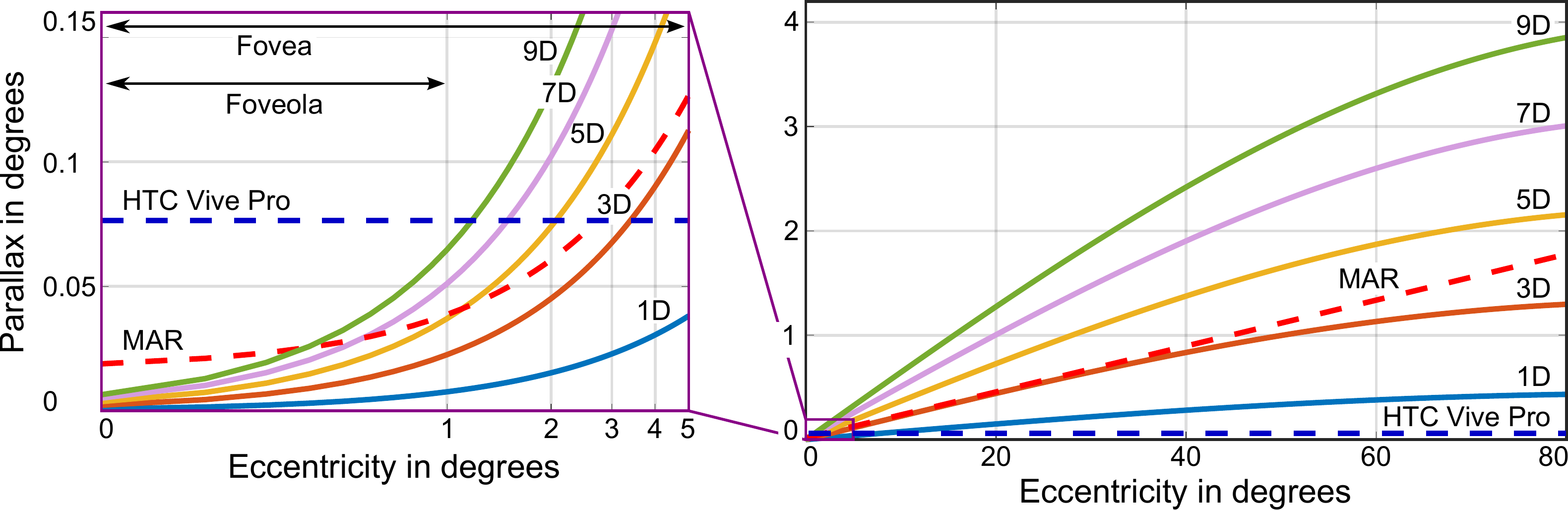}
		\caption{The amount of ocular parallax, measured in degrees of visual angle, increases with increasing eccentricity and relative distance between objects. Relative object distances of 3 diopters (inverse meters) and greater are above the minimum angle of resolution (MAR, red dashed line) and may therefore be detectable by a human observer. However, the amount of parallax on the foveola may be too small to be detected for static stimuli (left). The resolution available by modern VR displays, like the HTC Vive Pro (dashed blue lines), is slightly lower than the MAR of human vision in the foveola, but it exceeds the MAR for 2--3\textdegree\ of visual angle---precisely where we expect to see the effects of ocular parallax.}
		\label{fig:parallaxvsacuity}
\end{figure}

Figure~\ref{fig:parallaxvsacuity} illustrates the tradeoff between ocular parallax and visual acuity. Here, the minimum angle of resolution (MAR) represents a measure for visual acuity and approximates the resolution of human vision for a \emph{static} target. To model falloff of acuity in the peripheral visual field, we use the linear model proposed by Guenter et al.~\shortcite{Guenter:2012}: $\omega = m \, e + \omega_0$. Here, $\omega$ is the minimum angle of resolution (dashed red lines in Fig.~\ref{fig:parallaxvsacuity}), $e$ is the eccentricity in degrees, $m$ is the slope modeling the falloff of acuity, and $\omega_0$ is the MAR at the fovea. We set $\omega_0 = 1/60$ to model 20/20 vision and $m=0.022$ as proposed by Guenter et al.~\shortcite{Guenter:2012}. As seen in Figure~\ref{fig:parallaxvsacuity}, a relative distance of 3~D (diopters or inverse meters) should theoretically be detectable by a human observer for eccentricity angles smaller or equal to 40\textdegree. 

The left part of Figure~\ref{fig:parallaxvsacuity} also shows a semi-logarithmic plot zooming into the foveal region. We see that the magnitude of ocular parallax expected in the foveola, i.e. $e<1\degree$, may not be sufficient to be perceivable. Yet for eccentricities larger than 1\textdegree, relative object distances of 2--3~D may make ocular parallax a useful depth cue. The dashed blue lines in Figure~\ref{fig:parallaxvsacuity} also show the resolution of the HTC Vive Pro, one of the highest-resolution VR displays available today. The size of one pixel of this display is approximately 4.58~arcmin of visual angle, which is about $5\times$ higher than the MAR in the foveola. The resolution of this display starts to exceed the MAR at a distance of 2--3\textdegree\ of visual angle, implying that this technology is well suited to render ocular parallax precisely where it is expected to be perceived by the user.

However, ocular parallax is a motion cue and visual acuity alone may be insufficient to fully describe its perceived effects, because that model is only valid for static scenes. During eye movement, the perceived depth-dependent motion created by parallax results in time-varying retinal stimuli. Detection thresholds for differential velocities of this type have been shown to be considerably lower than the limits of visual acuity for all retinal loci. For example, ~\citet{Mckee:84} measured resolution thresholds of 2.7~arcmin and 4.8~arcmin in two participants at 10\textdegree\ eccentricity, but found that their comparable motion thresholds were less than 1~arcmin, indicating that the visual acuity-based analysis above is an overly conservative estimate for the conditions in which ocular parallax may be detectable.

\paragraph{\bf Retinal Blur}

As discussed above, larger relative distances between objects result in an increasing amount of ocular parallax. For vision in the physical world, such increasing distances also result in an increasing amount of defocus blur because objects at different optical distances require the eye to accommodate at one of them, placing the other out of focus. However, current VR/AR displays provide a substantially different viewing condition in that they usually provide only a single focal plane at a fixed optical distance. Users must accommodate to this fixed distance to perceive a sharp image~\cite{Padmanaban2183}. With AR/VR displays, we have full control over how much blur to render into the presented images and may choose to ignore this rendering altogether as it adds substantial computational cost. While depth-of-field rendering can be used to reduce visual discomfort in conventional stereo displays~\cite{Duchowski:2014}, it has not proven successful in driving accommodation or mitigating the vergence-accommodation conflict~\cite{mauderer2014,Konrad:2015,Johnson:16}, or acting as a reliable depth cue for single plane displays ~\cite{Marshall:96,Mather:02,Palmer:08,Zannoli:16}. Without depth-of-field rendering, ocular parallax is not affected by defocus blur in VR/AR, but it \textit{is} still affected by peripheral falloffs in acuity and differential motion thresholds. In the following section, we model and render ocular parallax without retinal blur as a computationally efficient approximation.

\section{Rendering Ocular Parallax}
\label{sec:ocular_parallax_rendering}

\subsection{Gaze-contingent Rendering}


In this section, we describe necessary modifications of the graphics pipeline to render ocular parallax. 

\paragraph{\bf Nodal Points}

\begin{figure}
	\centering
		\includegraphics[width=\columnwidth]{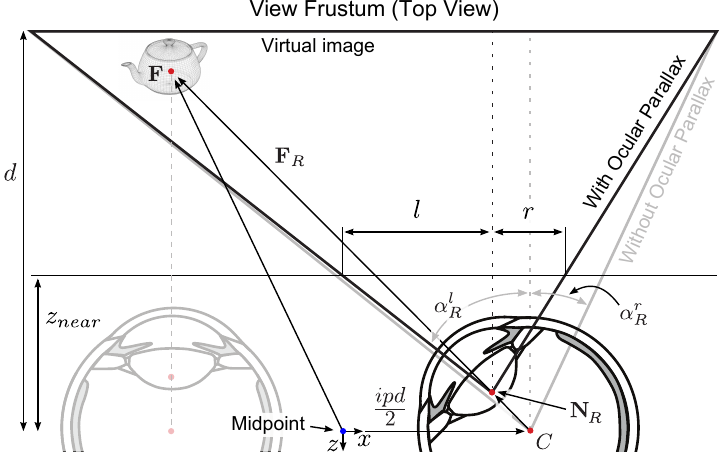}
		\caption{Illustration of parameters used to compute the nodal points $\mathbf{N}$ for each eye, defined with respect to the center of rotation $\mathbf{C}$ of the respective eye, from the fixation point $\mathbf{F}$, which is estimated by the eye tracking. The precise locations of these nodal points are required for calculating the view and projection matrices in the rendering pipeline.}
		\label{fig:projection}
\end{figure}

Ocular parallax is a gaze-contingent effect, and as such, eye tracking is necessary to render it appropriately. We assume that a binocular eye tracker estimates the 3D fixation point $\mathbf{F}$, which is typically defined with respect to the midpoint between the two eyes (see Fig.~\ref{fig:projection}). The center of rotation of each eye is offset from this midpoint by half the interpupillary distance (ipd). Defining the nodal points of each eye $\mathbf{N}_{L/R}$ relative to their respective center of rotation, they can be computed as
\begin{equation}
\mathbf{F}_{L/R} = \mathbf{F} \pm \left(\begin{smallmatrix} \frac{ipd}{2} \\ 0 \\ 0 \end{smallmatrix}\right), \quad \quad \mathbf{N}_{L/R} = \frac{NC}{|\mathbf{F}_{L/R}|}\mathbf{F}_{L/R}
\end{equation}
where $\mathbf{F}_{L/R}$ defines the fixation point relative to each eye's center of rotation $\mathbf{C}$ and $NC$ is the distance between the center of rotation and the front nodal point. The locations of these nodal points are then used to update the view and projection transforms in each rendered frame.

\paragraph{\bf View and Eye Matrix}

The standard graphics pipeline transforms each vertex $\mathbf{v}$ to view space by multiplying it with the model ($\mathbf{M}$) and view ($\mathbf{V}$) matrices. In binocular displays, such as VR and AR systems, an additional per-eye translation by half the ipd is applied to create correct stereoscopic cues by transforming vertices into eye space. To account for ocular parallax, we modify the transform to eye space with an additional translation by $-\mathbf{N}_{L/R}$. The full transformation from each vertex to eye space is then defined as
\begin{align}
	\mathbf{v}_{L/R}^{(eye)} & = \mathbf{E}_{L/R} \cdot \mathbf{V} \cdot \mathbf{M} \cdot \mathbf{v}, \label{eq:vertex_to_eye} \\
	\mathbf{E}_{L/R} & =
	\begin{bmatrix}
    1 & 0 & 0 & -N_{L/R}^{(x)} \\
		0 & 1 & 0 & -N_{L/R}^{(y)} \\
		0 & 0 & 1 & -N_{L/R}^{(z)} \\
		0 & 0 & 0 & 1
	\end{bmatrix}
	\begin{bmatrix}
    1 & 0 & 0 & \pm \frac{ipd}{2} \\
		0 & 1 & 0 & 0 \\
		0 & 0 & 1 & 0 \\
		0 & 0 & 0 & 1
	\end{bmatrix},
\end{align}
where $\mathbf{E}_{L/R}$ is the eye matrix, i.e. the transformation from view space to eye space, and $\mathbf{v}_{L/R}^{(eye)}$ defines each vertex in eye space.

\paragraph{\bf Projection Matrix}

Vertices in eye space are transformed into clip space using the projection matrix. A perspective projection in stereo rendering is usually represented as an asymmetric off-axis view frustum defined by a near and far clipping plane, $z_{near}$ and $z_{far}$, as well as the left ($l$), right ($r$), top ($t$) and bottom ($b$) boundary values on the near clipping plane~\cite{Shirley:2009}. Using a right-handed coordinate system, the corresponding projection matrix has the general structure outlined in \autoref{eq:projMat}. For clarity, we only show the projection transform of the right eye, but a matrix of similar form is applied for the left eye:
\begin{align}
	\mathbf{v}_{L/R}^{(clip)} &= \mathbf{P}_{L/R} \cdot \mathbf{v}_{L/R}^{(eye)}, \label{eq:projOp}\\
	\mathbf{P}_{R} = &
	\begin{bmatrix}
    \frac{2 \cdot z_{near}}{{r_{R} - l_{R}}} & 0 & \frac{{r_{R} + l_{R}}}{{r_{R} - l_{R}}} & 0 \\
		0 & \frac{2 \cdot z_{near}}{{t_{R} - b_{R}}} & \frac{{t_{R} + b_{R}}}{{t_{R} - b_{R}}} & 0 \\
		0 & 0 & \frac{-(z_{far} + z_{near})}{z_{far} - z_{near}} & \frac{-2 \cdot z_{far} \cdot z_{near}}{z_{far} - z_{near}} \\
		0 & 0 & -1 & 0
	\end{bmatrix}.	 \label{eq:projMat}
\end{align}
The frustum boundary values are determined from parameters of the physical setup, including the field of view of the display, the distance to the virtual image\footnote{We make the assumption that the virtual image is at optical infinity, i.e. $d=\infty$, such that distant objects do not move during eye motion but objects at closer distances shift relative to the background.} $d$ and the position of the front nodal points $\mathbf{N}_{L/R}$ defined with respect to the corresponding center of rotation $\mathbf{C}$. Assuming that we know the fields of view $\alpha^{\{l,r,t,b\}}$ defining the asymmetric frustum of the conventional stereo rendering mode (see Fig.~\ref{fig:projection}), we can compute the asymmetric view frustum of ocular parallax rendering as 
\begin{align}
	\{l_{R}, r_{R}\} & = \frac{z_{near} + N_{R}^{(z)}}{d + N_{R}^{(z)}} \left( d \cdot \textrm{tan} \left(\alpha_{R}^{\{l,r\}} \right) + N_{R}^{(x)} \right), \label{eq:frustumLR}\\
	\{t_{R}, b_{R}\} & = \frac{z_{near} + N_{R}^{(z)}}{d + N_{R}^{(z)}} \left( d \cdot \textrm{tan} \left(\alpha_{R}^{\{t,b\}} \right) + N_{R}^{(y)} \right). \label{eq:frustumTB}
\end{align}
The projection matrix is updated on a per-frame basis using the tracked nodal points. Applying the above modifications to the view and projection transforms renders perceptually accurate ocular parallax using slight modifications of the graphics pipeline, under the assumption that there are no optical distortions. Currently, we use the manufacturer-supplied optical distortion correction to account for optical aberrations of the HMD lenses across the visual field.

\subsection{Perceptual Effects of Ocular Parallax Rendering}
\label{sec:implementation:perceptual_effects}

Ocular parallax rendering is expected to have implications on several depth cues. We briefly discuss these here to motivate the user experiments that specifically evaluate each of these effects in the following sections.

\paragraph{\bf Micro Parallax}

Parallax describes both magnitude and direction of the retinal velocity of objects at different depths while either the scene is in motion or the user's head moves. We expect a similar effect to occur, albeit at a significantly smaller magnitude, due to changes in gaze direction when ocular parallax rendering is enabled. Such changes in gaze direction rotate the nodal point about the center of rotation, inducing tiny amounts of depth-dependent ``micro parallax'' into the retinal image. 
The relative magnitudes of the velocities of objects at different depths could provide an ordinal depth cue, helping users better understand the relative depth ordering of a scene. This has, for example, been shown to be the case for head motion parallax~\cite{Yonas:87}, but it is not clear whether ocular parallax is an effective ordinal depth cue as well. Furthermore, the absolute magnitude of retinal velocity induced by ocular parallax could serve as an absolute depth cue, but the small retinal velocity magnitudes may not be sufficient to robustly estimate absolute distance. While micro parallax has an effect on all conducted user experiments, we specifically quantify its perceptual effect size in the discrimination threshold experiment in Section~\ref{sec:thresholds:discrimination}.



\paragraph{\bf Gaze-contingent Occlusion}

Micro parallax near occlusion bou\-ndaries is particularly interesting because there we observe gaze-contingent occlusion (see Fig.~\ref{fig:teaser}). When objects at different depths overlap, the observed parallax due to eye rotations causes the accr\-etion-deletion of only the farther object's texture. While occlusion can only provide ordinal depth information, it is considered one of the strongest depth cues in static environments~\cite{Cutting:1995}. Particularly relevant to gaze-contingent occlusion is the fact that time-varying accretion-deletion of texture due to head-motion-induced parallax has been shown to be an effective ordinal depth cue~\cite{Yonas:87}. Yet, it is unknown whether the same is true for the small amounts of accretion-deletion of texture observed with ocular parallax rendering. We evaluate the perceptual effect size of gaze-induced occlusion by estimating the detection thresholds for ocular parallax in Section~\ref{sec:thresholds:detection} and its effectiveness as an ordinal depth cue in Section~\ref{sec:depth:ordinal}.


\paragraph{\bf Disparity Distortion}

Conventional stereoscopic rendering techniques assume that the centers of projection and rotation of the eyes are equivalent. We show that this is not the case. Therefore, ocular parallax could, in principle, also affect the rendered disparity values in stereographic image pairs and therefore distort perceived depth. We evaluate this hypothesis with a user experiment that studies the effect of ocular parallax on absolute depth perception in Section~\ref{sec:depth:metric}.

\subsection{Implementation}
\label{sec:implementation}





\paragraph{\bf Hardware} We implement ocular parallax rendering with a prototype virtual reality system. The VR system is an HTC Vive Pro connected to the open source binocular Pupil Labs eye tracker. The HTC Vive Pro has a field of view of 110\textdegree, a refresh rate of 90~Hz, and a $1440\times1600$ pixel organic light-emitting diode display, resulting in a theoretical resolution of 4.58~arcmin/pixel. The HTC Vive Pro supports built-in ipd adjustment. The Pupil Labs eye tracker snaps into the the HTC Vive Pro and estimates a global fixation point (i.e. $\mathbf{F}$) at 120~Hz, supporting about 1\textdegree\ of manufacturer-reported gaze accuracy and 0.08\textdegree\ of gaze precision (see supplement for additional characterization). In the egocentric depth perception study, an HTC Vive Tracker estimates the position and orientation of the users' hand at 120~Hz with an accuracy of 1.9~mm root-mean-square-error (RMSE) and a precision of 1.5~mm RMSE\footnote{\url{http://doc-ok.org/?p=1478}}. 

\paragraph{\bf Software and Calibration} Unity was used as the rendering engine for both the ocular parallax rendering and the user experiments. Pixel-precise rendering and anti-aliasing were used for all rendered stimuli. Pupil Labs provides a Unity plugin that interfaces with their Python library, providing eye-tracking calibration and gaze tracking. All software related to acquiring subject data was written as C\# scripts in Unity. The data were then analyzed in Python.

\section{Perceptual Thresholds for Ocular Parallax in VR}
\label{sec:thresholds}

\begin{figure*}[t]
	\centering
		\includegraphics[width=\textwidth]{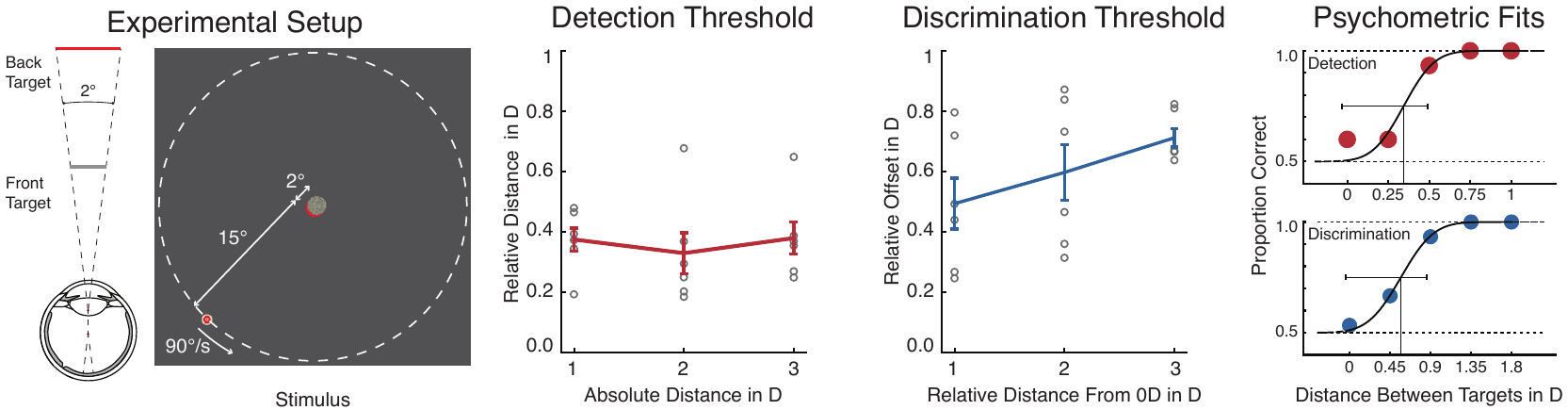}
		\caption{Detection and discrimination thresholds for ocular parallax in VR. Two experiments were conducted to estimate perceptual thresholds using an HTC Vive Pro head mounted display presenting stimuli of the form shown on the left. A red surface subtending 2\textdegree\ of visual angle was completely occluded by a noisy gray surface in front of it (left). The relative distances of these surfaces were varied with conditions described in the text. Detection thresholds (center left) and discrimination thresholds (center right) were estimated from the psychometric functions fitted to the recorded subject data (examples shown right).}
		\label{fig:thresholds}
\end{figure*}


The primary goal of this section is to establish depth-dependent detection and discrimination thresholds for ocular parallax in VR. Although ocular parallax has been shown to be well within the range of human visual acuity for natural viewing conditions (e.g.,~\citet{Bingham:1993}), we are not aware of any work that actually verified that this small effect size is even perceivable with the limited resolution offered by current-generation VR displays or, if it is, what the specific thresholds are. These thresholds are crucial for understanding in which conditions ocular parallax rendering is visible and how reliable of a depth cue it may be. To estimate these thresholds, we perform two psychophysical experiments that are discussed in the following sections with the apparatus described in Section~\ref{sec:implementation}.

\subsection{Detection Thresholds for Ocular Parallax}
\label{sec:thresholds:detection}

With this experiment, we aim to estimate a depth-dependent detection threshold at which ocular parallax is perceivable. 

\paragraph{\bf Stimuli} As seen in Figure~\ref{fig:thresholds} (left), we presented two circular surfaces at varying distances to the subject. The surfaces were scaled to subtend 2\textdegree\ of visual angle irrespective of depth. The farther surface was solid red and the front one was textured with white noise. Without ocular parallax enabled, the front surface exactly occluded the back one.

An additional small red and white fixation target was then rendered, circling around the scene at a distance of 16\textdegree\ from the center, or 15\textdegree\ from the edges of the front and back surfaces. This was the largest eccentricity that kept the target within the aberration-free viewing zone of the Vive Pro. This fixation target rotated with an angular velocity of $90\degree/\si{\second}$, resulting in a perceived retinal motion of $24.81\degree/\si{\second}$, which falls within smooth pursuit rates~\cite{Westheimer:54}. The starting position and rotation direction of the fixation target were randomized per trial.

\paragraph{\bf Conditions} All stimuli were presented monocularly to the right eye. In each trial, the absolute distance between the viewer and back surface was randomly chosen as 1, 2, or 3~D. The relative distance between the front and back surface was also randomly chosen as 0, 0.25, 0.5, 0.75, or 1~D. In each trial, subjects viewed the stimulus with ocular parallax rendering enabled and disabled. These two conditions were presented in different 2-second intervals in random order, separated by a 0.5~second blank frame. 

\paragraph{\bf Subjects} Six adults participated (age range 26--38, 1 female). Due to the demanding nature of our psychophysical experiment, only a few subjects were recruited, which is common for low-level psychophysics (see e.g.~\citeN{Patney:2016}).

All subjects in this and all following experiments had normal or corrected to normal vision, no history of visual deficiency, and no color blindness. All subjects gave informed consent. The research protocol was approved by the Institutional Review Board at the participating university. 

\paragraph{\bf Procedure}

To start the session, each subject performed a 7-point eye-tracker calibration that is provided by the manufacturer. To minimize eye-tracker error over the course of the experiment, each trial began with a single-point re-calibration of the eye trackers; subjects were instructed to fixate on a cross target centered on the screen, and the measured gaze direction was used to compensate for possible drift error. Subjects then viewed the stimulus rendered with one of the two ocular parallax rendering conditions for 2 seconds, then a blank screen for 0.5 seconds, and then again the stimulus with the other rendering condition from the first interval for another 2 seconds. Each trial constituted a two-alternative forced choice test, and subjects were asked to choose the time interval which exhibited more relative motion between the two surfaces with a keyboard. This concluded the trial. No feedback was provided. Subjects were instructed to fixate only on the moving fixation target, and never to the center surfaces. 

There were 15 distance configurations and 15 trials for each configuration for a total of 225 trials per subject. The experiment took about 25 minutes to complete, including instruction and eye tracking calibration per subject.

\paragraph{\bf Analysis}

For each of the 15 distance configurations, we computed the proportion of correct responses. Using Bayesian inference methods~\cite{Shutt:06,Wichmann:2001a,Wichmann:2001b}, we fit a psychometric function to each subject's performance at each of the three absolute depths of the back surface. Each psychometric function gives us a detection threshold, measured in diopters of relative distance from the absolute distance of the back surface. The thresholds represent where the psychometric function exceeded a 75\% chance for a correct response. An example of one of these psychometric functions is shown in Figure~\ref{fig:thresholds} (top right) and all measured psychometric functions are available in the supplement.

\paragraph{\bf Results} The detection thresholds, averaged across subjects, are plotted in \autoref{fig:thresholds} (center left). We see that these thresholds are invariant to the absolute distance of the back surface. This is expected because the conditions were equally spaced in diopters and ocular parallax, like other types of parallax cues, is perceptually linear in dioptric space (see supplement for an analysis). The estimated detection threshold is approximately 0.36~D. This implies that, for an eccentricity of 15\textdegree\ and a relative object distance as low as 0.36~D, ocular parallax may be perceivable in VR. 

This result is surprising because visual acuity alone (see Fig.~\ref{fig:parallaxvsacuity}) predicts detection thresholds that are an order of magnitude higher that what we measured. Yet, our results are consistent with data reported for physical viewing conditions (i.e. non-VR settings)~\cite{Bingham:1993} and with differential velocity thresholds~\cite{Mckee:84}. We expect the detection thresholds to increase linearly with eccentricity in the extra-foveal region because both the magnitude of micro parallax (see Fig.~\ref{fig:parallaxvsacuity}) and differential velocity thresholds increase roughly linearly there~\cite{Mckee:84}. Also, we expect detection thresholds to decrease with faster eye movements because differential velocity thresholds inversely vary with retinal velocity~\cite{Mckee:84}.

Our results emphasize that the human visual system is much more sensitive to small amounts of motion, even in peripheral vision, than na\"ively expected. Even the small amount of parallax induced by ocular parallax rendering may be visible in many VR applications. An important question that arises from this insight is whether ocular parallax rendering can improve depth perception or the realism of a 3D scene. We perform several experiments in Sections \ref{sec:depth} and \ref{sec:realism} that aim at answering this question.

\subsection{Discrimination Thresholds for Ocular Parallax}
\label{sec:thresholds:discrimination}

A discrimination threshold tells us what the smallest amount of perceivable change in ocular parallax is. For a fixed eye eccentricity, this threshold depends on both the absolute distance of the reference surface (i.e., the back surface) and also the relative offset from that surface. Conceptually, one would have to estimate discrimination thresholds for each combination of absolute and relative distance, which would be an arduous task. However, due to the fact that the detection thresholds are depth independent, we assume that the discrimination thresholds are also independent of the absolute distance to the back surface. This assumption makes it easier to set up an experiment to estimate discrimination thresholds for ocular parallax in VR, which we did by performing a second experiment that uses the same apparatus, stimuli, and analysis as the first experiment and a very similar procedure, but with slightly different conditions. Six adults participated (age range 26--32, 1 female). 

\paragraph{\bf Conditions} The back surface was fixed to 0~D for all conditions. Instead of presenting the stimulus with ocular parallax rendering enabled for only one interval, as done in the previous experiment, we enabled ocular parallax rendering for both intervals. The relative offset from the back surface was randomly chosen as 1, 2, or 3~D and assigned as the depth for one of the front surfaces shown in each trial. The other appeared at one of the following distances from the previously chosen relative offset: 0, 0.45, 0.9, 1.35, and 1.8~D for the 1~D and 2~D initial offsets; and 0, 0.7, 1.4, 2.1, and 2.8~D for the 3~D initial offset. Again, there were 15 distance configurations and 15 trials for each configuration for a total of 225 trials per subject.

\paragraph{\bf Results} We used a similar analysis as the one described in the previous subsection to estimate the discrimination thresholds, which are plotted in \autoref{fig:thresholds} (center right). As expected, the discrimination thresholds increase linearly with increasing amounts of ocular parallax, per Weber's law, but due to the proximity to the detection threshold magnitude, the slope is less than 1. A linear fit between the discrimination thresholds and relative offset from 0~D has a slope of 0.11 and an intercept of 0.38~D, further verifying the measured 0.36~D detection threshold from the previous experiment.

In conclusion, the estimated discrimination thresholds are well within the range of the depth ranges of natural scenes commonly rendered for immersive virtual reality applications. This result motivates further studies to investigate if or when ocular parallax could be used as a reliable depth cue in these applications. We take a first step at answering this question by conducting experiments that study whether ocular parallax has a measurable effect on depth perception in the following sections.

\section{Ocular Parallax and Depth Perception}
\label{sec:depth}


Motivated by the discussion in Section~\ref{sec:implementation:perceptual_effects} and the surprisingly low detection and discrimination thresholds measured in the previous section, we proceed to investigate the importance of ocular parallax as a depth cue. Here, we distinguish between ordinal depth perception, which provides information about the \emph{relative} depth ordering of a scene (i.e., object A is closer than object B), and metrical depth cues, which also provide \emph{absolute} distance estimates (i.e., object A is 1~m away and object B is 2~m away). It seems intuitive that gaze-contingent occlusion can provide ordinal depth information due to the deletion and accretion of the farther object's texture near occlusion boundaries. Moreover, gaze-induced micro parallax also provides ordinal information in the relative magnitudes of retinal velocities of objects at different depths. However, the effectiveness of ocular parallax as an absolute depth cue is questionable because detecting absolute motion velocities of objects at different depths may be unreliable, and occlusion only provides ordinal depth information. We investigate the effect of ocular parallax rendering on both ordinal and absolute depth perception in the following experiments with the apparatus described in Section~\ref{sec:implementation}.


\subsection{Effect of Ocular Parallax on Ordinal Depth Estimation}
\label{sec:depth:ordinal}

In this section, we study the benefits of ocular parallax rendering on ordinal depth estimation. We also  investigate whether depth perception gained from ocular parallax is purely a visual process, or requires some non-visual, extra-retinal signal like the magnitude or direction of eye rotation. The perception of depth from head-motion parallax, for example, has been shown to rely on an extra-retinal signal in the form of optokinetic response eye movements~\cite{Nawrot:03}. This experiment is modeled after a related experiment investigating retinal blur as an ordinal depth cue~\cite{Zannoli:16}. Twenty-one adults participated (age range 22--43, 4 females), of which two were excluded for failing to follow instructions.

\begin{figure}[t]
	\centering
		\includegraphics[width=\columnwidth]{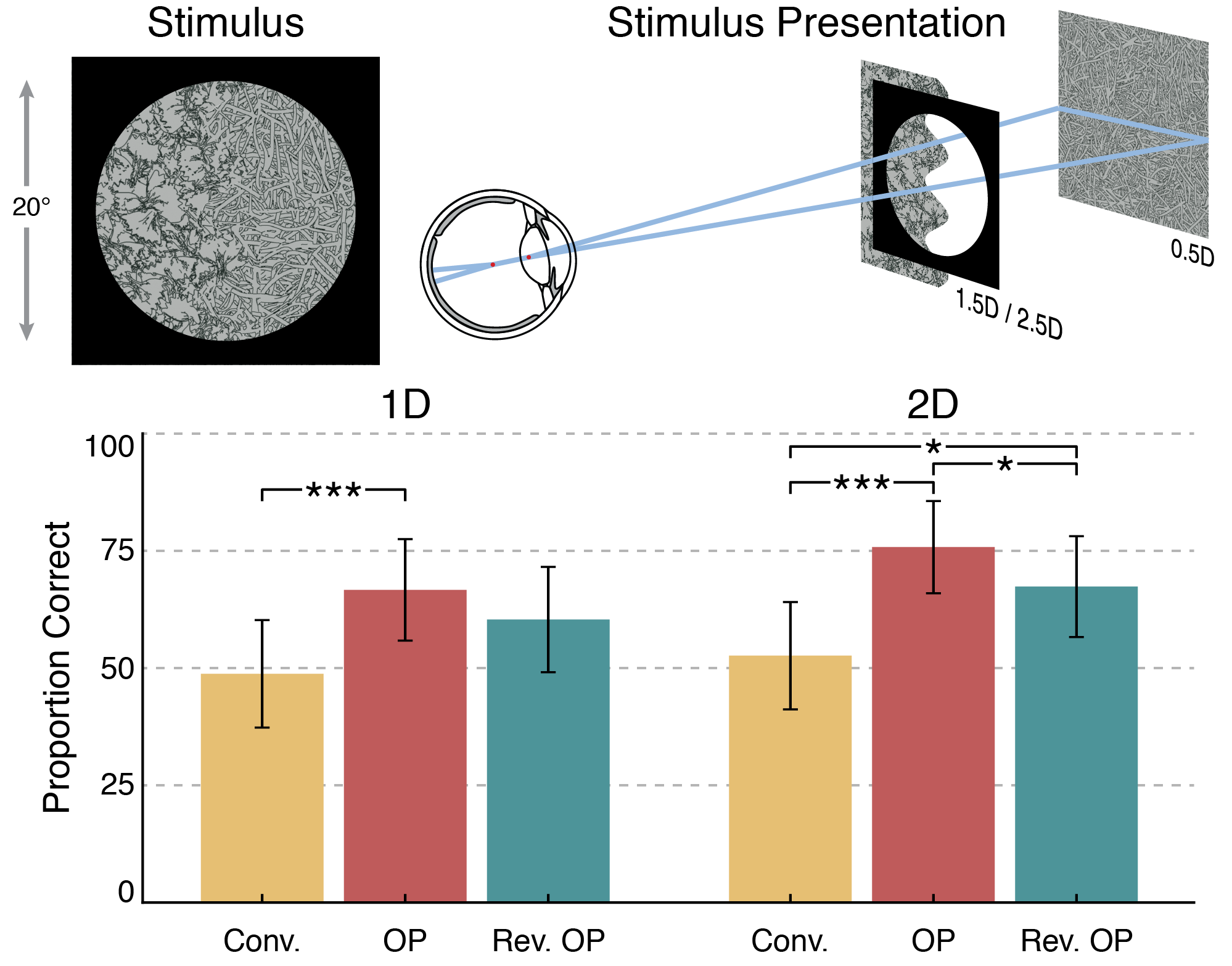}
		\caption{Effect of ocular parallax on ordinal depth perception.	Subjects viewed the monocular stimuli consisting of two distinctly textured surfaces separated by 1~D or 2~D (top) and were asked which one was closer. The proportions of correct responses, averaged across subjects per condition, are plotted on the bottom. Subjects performed significantly better with ocular parallax rendering enabled compared to conventional rendering. However, they also performed slightly better than conventional rendering with reversed ocular parallax rendering, indicating that the extra-retinal signal of eye rotation may not be crucial for depth perception. Significance is indicated at the $p\leq0.05$, 0.01, and 0.001 levels with \textasteriskcentered, \textasteriskcentered\textasteriskcentered, and \textasteriskcentered\textasteriskcentered\textasteriskcentered, respectively. Error bars represent standard error.}
		\label{fig:ordinal}
\end{figure}

\paragraph{\bf Stimuli} The monocular stimuli, displayed on a single virtual image plane, consisted of two differently textured, frontoparallel surfaces at different depths (Fig.~\ref{fig:ordinal}, top). The textures, the same as those in the aforementioned depth order study~\cite{Zannoli:16}, had the same space-average luminance and contrast energy, and exhibited similar amplitude spectra. The rear surface was fixed at 0.5~D and the front surface appeared at either 1.5~D or 2.5~D. The border between the two surfaces had a sinusoidal shape. The surfaces were scaled to subtend the same visual angle irrespective of depth. Subjects viewed the surfaces through a 20\textdegree\ circular aperture that was unaffected by ocular parallax rendering. 


\paragraph{\bf Procedure} First, all subjects performed a 7-point eye-tracker calibration provided by the manufacturer to start the session. As in the psychophysical experiments, each individual trial began with a single-point re-calibration of the eye trackers. Following the re-calibration, the stimulus was presented monocularly to the right eye for 3 seconds with one of the three rendering conditions. Subjects were instructed to freely gaze anywhere within the aperture where both surfaces were visible. The stimulus was then replaced with a prompt asking which surface, left or right, appeared nearer, and a keyboard response concluded the trial. No feedback was provided.

\paragraph{\bf Conditions} We evaluated three different rendering conditions (conventional, ocular parallax, and reversed ocular parallax) and two near surface distances (1.5~D and 2.5~D). With the monocular stimuli devoid of perspective and accommodation/retinal blur cues, the only expected sources of depth information are gaze-contingent micro parallax and occlusion. We included both correct and reversed ocular parallax rendering to understand whether depth perception in this scenario is purely a visual process, requiring only the retinal image with micro parallax, or whether it also requires an extra-retinal signal in the form of eye rotation direction. With reversed ocular parallax rendering, the sign of the visual signal is negated, causing occluded surface textures to accrete when they would normally be deleted. Subjects expecting the visual and extra-retinal signals to be consistent could misinterpret depth orderings. Reversed ocular parallax was implemented by negating the $x$ and $y$ components of the estimated nodal points of the eyes, $\mathbf{N}_{L/R}$.

The near surface distances create a 1~D and 2~D separation from the rear surface, resulting in different magnitudes of the gaze-contingent effects. Following the threshold experiments, we expect the proportion of correct responses to increase with surface separation. Overall, there were 6 conditions and each condition was evaluated with 15 trials for a total of 90 trials per subject. 

\paragraph{\bf Results} The proportion of correct responses, averaged across subjects per condition, is plotted in \autoref{fig:ordinal} (bottom). As expected, subjects in the conventional static rendering condition performed close to random, correctly identifying the nearer surface in 48.8\% and 52.6\% of trials for 1~D and 2~D of surface separation, respectively. Enabling ocular parallax rendering clearly improved ordinal depth judgment with subjects performing at 66.7\% and 75.8\% correct identification for the 1~D and 2~D separations, respectively. Subjects in the reversed ocular parallax condition fell in-between, performing at 60.4\% and 67.4\% correct identification for the two separation distances.

We conducted a 2$\times$3 repeated-measures ANOVA on the proportion of correct responses with independent variables of rendering mode (conventional, ocular parallax, reversed ocular parallax) and separation distance (1~D or 2~D). Greenhouse-Geisser sphericity correction was applied. The ANOVA shows a very significant effect of rendering mode ($F(1.7, 30.65) = 17.98$, $p < 0.0001$) as well as a significant effect of distance ($F(1, 18) = 8.14$, $p < 0.05$). The ANOVA does not reveal a significant interaction between rendering mode and distance ($F(1.78, 31.96) = 0.42$, $p = 0.64$).

Post-hoc tests were conducted as pairwise $t$-tests between rendering modes at each separation distance, with Bonferroni correction applied to the $p$-values. The post-hoc tests found that, at 1~D of separation, ocular parallax rendering shows a significant improvement over conventional rendering ($p<0.001$), but not over reversed ocular parallax rendering. Reversed ocular parallax rendering does not show a significant improvement over conventional rendering. At 2~D of separation, ocular parallax rendering shows a significant improvement over conventional rendering ($p<0.001$) as well as reversed ocular parallax rendering ($p<0.05$). Reversed ocular parallax rendering also shows a significant improvement over conventional rendering ($p<0.05$) at this separation.

In summary, this experiment demonstrates that ocular parallax significantly improves ordinal depth perception over conventional rendering. However, reversed ocular parallax also improves ordinal depth perception over conventional rendering, but not nearly as much as when rendered correctly. The reduced performance in the reversed ocular parallax condition compared to the correct ocular parallax condition suggests that extra-retinal signals, like eye rotation, play an important role in the perception of depth. However, this effect seems to be weaker than for motion parallax, where a reversal in eye movement causes a reversal in the sign of the perceived depth~\cite{Nawrot:03}. Still, subjects performed better, in some conditions significantly so, when the directions of the retinal image motion and eye rotation were consistent. Therefore, enabling correct ocular parallax rendering can benefit ordinal depth estimation which can be particularly useful for viewing 3D scenes that often contain many occlusion boundaries.

\subsection{Effect of Ocular Parallax on Absolute Depth Estimation}
\label{sec:depth:metric}

\begin{figure}[t]
	\centering
		\includegraphics[width=\columnwidth]{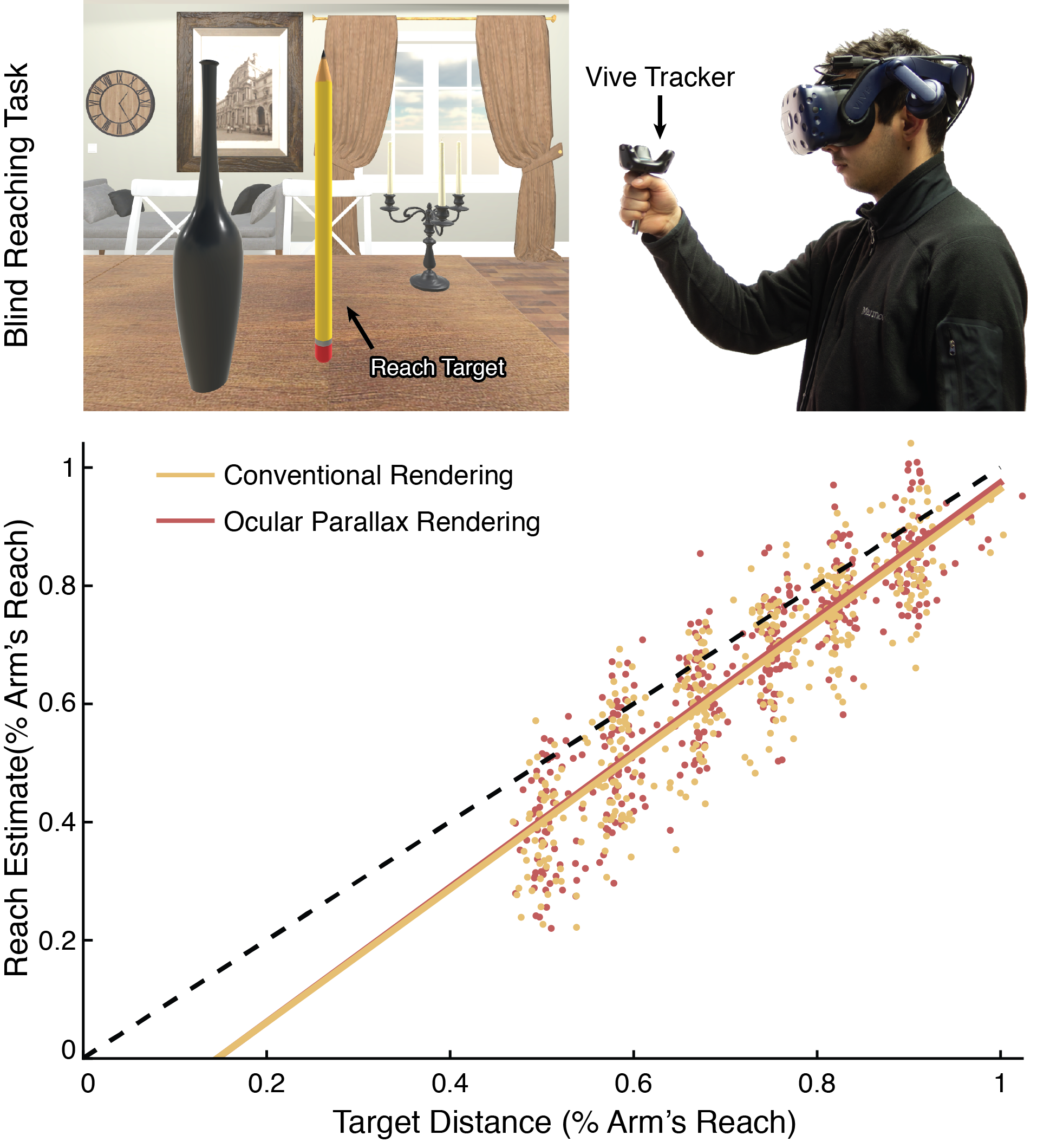}
		\caption{
		Effect of ocular parallax on absolute, egocentric depth perception. Subjects viewed a target (pencil) and then, with the display turned off, were asked to reach for it (top right). We did not find a statistically significant difference between conventional stereo rendering and stereo rendering with ocular parallax in this experiment (bottom plot).}
		\label{fig:egocentric}
		\vspace{-14pt}
\end{figure}

Next, we study whether ocular parallax rendering can benefit egocentric distance (distance from one's self) estimation in a stereo environment. While unlikely, gaze-induced micro parallax and the reduction in peripheral disparity errors could affect absolute distance estimation. 

In the experiment, subjects performed a blind reaching task introduced by ~\citet{Napieralski:11} to estimate distances to objects because verbal estimates have been shown to be inaccurate~\cite{Renner:2013}. In a photo-realistic dining room scene, subjects viewed a stereoscopically rendered pencil for a minimum of 5 seconds after which the screen was blanked and they reached their hand to where they had last seen the pencil (\autoref{fig:egocentric}, top left). Each subject performed the task with ocular parallax enabled and disabled, and the reach target distances were set proportionally \Dash 50, 58, 67, 75, 82, and 90\% \Dash to each subjects' maximum arm reach. Each of the 12 conditions were evaluated with five blind reaching tasks for a total of 60 randomly-ordered trials per subject. During the session, 6-DOF head pose tracking was enabled, and the hand's position was tracked via an HTC Vive tracker mounted on an optical post (\autoref{fig:egocentric}, top right). Sixteen young adults participated (age range 22--32, 5 females), of which one was excluded for not passing a standard Randot stereo vision test, and two others were excluded due to the eye tracker failing to track their pupils. 


Subjects saw little to no improvement in their egocentric distance estimates with ocular parallax rendering enabled. Figure~\ref{fig:egocentric} shows each subject's reach, as a percentage of their maximum arm reach, to a presented target distance for the ocular parallax enabled and disabled conditions. Linear models were fit to the two sets of reaches; the slopes for the ocular parallax enabled and disabled modes were 1.141 and 1.128, respectively, while the intercepts were 0.167 and 0.165, respectively. A multiple regression analysis did not show a significant difference between the two conditions. It is therefore unlikely that the systematic underestimation of distances to virtual objects compared to real ones~\cite{Renner:2013} can be explained by the omission of ocular parallax rendering. For more details regarding this experiment, please refer to the supplement.

\section{Ocular Parallax and Perceptual Realism}
\label{sec:realism}

We next measured the effect of ocular parallax on perceptual realism. Twenty-one adults participated in the experiment (age range 22--43, 4 females), of which two were excluded for failing to follow instructions.

\begin{figure}[t]
	\centering
		\includegraphics[width=\columnwidth]{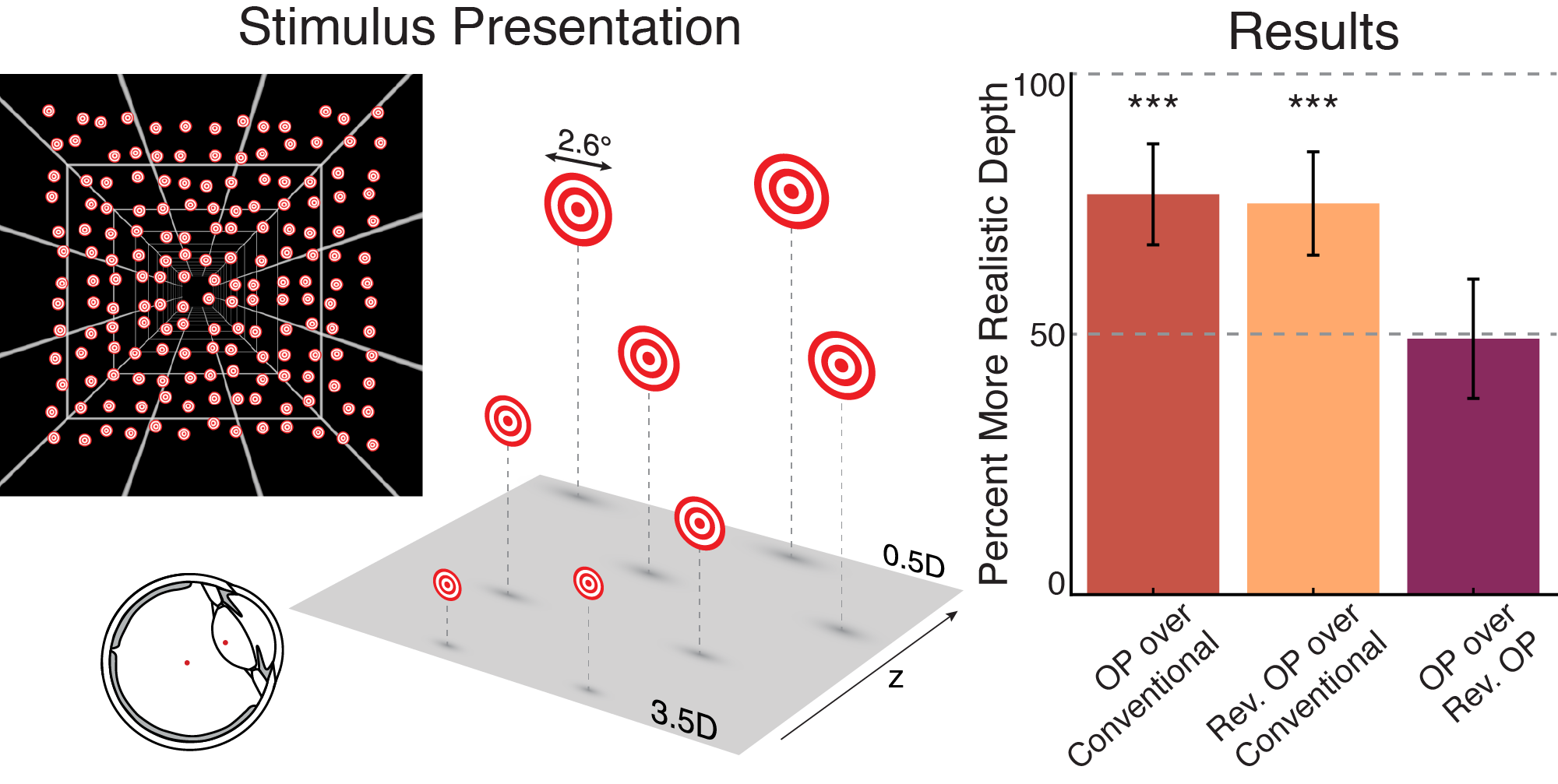}
		\caption{Evaluating perceptual realism. Subjects viewed a 3D scene consisting of targets that are randomly distributed in depth but that do not occlude one another (left). 
		This stimulus was presented with either conventional, ocular parallax, or reversed ocular parallax rendering and we asked subjects to indicated which rendering mode provided a stronger impression of realistic depth. Results of pairwise comparisons between these rendering modes show the percent of times the first member of the pair was chosen over the second (right). Rendering with correct and reversed ocular parallax conveyed a stronger impression of depth compared to conventional rendering, but when compared against one another no difference was observed. This result indicates that the relative magnitudes of depth-dependent motion velocities are most important for perceptual realism but not necessarily their direction. Significance is indicated at the $p\leq0.05$, 0.01, and 0.001 levels with \textasteriskcentered, \textasteriskcentered\textasteriskcentered, and \textasteriskcentered\textasteriskcentered\textasteriskcentered, respectively. Error bars represent standard error.}
		\label{fig:realistic_depth}
\end{figure}

\paragraph{\bf Stimuli and Conditions} A 3D scene~(Fig.~\ref{fig:realistic_depth}, left) was presented to the right eye with conventional, ocular parallax, or reversed ocular parallax rendering. The scene was composed of four grid planes extending into the distance and a field of frontoparallel targets. These targets were randomly distributed in depth between 0.5~D and 3.5~D and were scaled to subtend 2.6\textdegree\  of visual angle regardless of depth. The targets' lateral positions were randomly chosen such that they did not occlude one another and the entire target field subtended 60\textdegree. 

\paragraph{\bf Procedure} First, all subjects performed a 7-point eye-tracker calibration provided by the manufacturer to start the session. As in the psychophysical experiments, each individual trial began with a single-point re-calibration of the eye trackers. Then, the target positions used for the subsequent stimuli were generated. The first stimulus was presented for 3 seconds with one of the three rendering conditions. The targets then disappeared for 1 second, leaving only the grid pattern shown with conventional rendering. The second stimulus was then presented for 3 seconds using one of the three rendering conditions, but not the same as the one used for the first stimulus. After the second stimulus, a prompt appeared asking the subjects to determine, in a forced-choice judgment, the stimulus that portrayed a stronger impression of realistic depth. No feedback was provided. Each trial therefore consisted of three pairwise comparisons: \textit{conventional} vs. \textit{ocular parallax}, \textit{conventional} vs. \textit{reversed ocular parallax}, and \textit{ocular parallax} vs. \textit{reversed ocular parallax}. 

\paragraph{\bf Results} The results of the pairwise comparisons are averaged across users and trials and plotted in~\autoref{fig:realistic_depth} (right). In 76.8\% of trials, subjects reported a stronger sense of realistic depth when viewing ocular parallax rendering over conventional rendering. Interestingly, subjects also reported more realistic depth for reversed ocular parallax rendering over conventional rendering in 75.1\% of trials. Each of these percentages is significantly greater than 50\% ($p<0.001$ and $p<0.001$ respectively, one-tailed binomial test). Users had more difficulty differentiating between the two ocular parallax modes and reported more realistic depth when ocular parallax was correctly rendered in only 49.1\% of trials, which was not significantly greater than 50\%. 

These results suggest that gaze-induced micro parallax, offered by ocular parallax rendering, has a significant effect on perceptual realism when compared to conventional rendering. Moreover, the results suggest that the relative motion magnitudes of the depth-dependent retinal velocities are more important than their direction. This insight further emphasizes that, similar to the ordinal depth perception task (Sec.~\ref{sec:depth:ordinal}), extra-retinal signals were likely not notably factored into the perception of depth on this subjective task. Unlike the experiment in Section~\ref{sec:depth:ordinal}, where both gaze-contingent occlusion and micro parallax were present as cues, in this experiment, micro parallax was the primary indicator of depth because the targets did not overlap.

\section{Discussion}
\label{sec:discussion}

In summary, our primary contribution is to introduce a new technology for virtual and augmented reality: ocular parallax rendering. This technique is enabled by eye tracking systems which can already be found in some of the latest headsets, like the Microsoft Hololens~2, Magic Leap One, Varjo, Fove, and HTC Vive Pro Eye. Ocular parallax rendering could be jointly implemented with other gaze-contingent rendering methods, such as foveated rendering, and it requires no additional computational cost compared to conventional stereo rendering. 

To evaluate ocular parallax rendering, we designed and conducted a series of user experiments. First, we measured detection thresholds and show that the effect is perceivable when the relative distance between two partially occluding objects is as low as 0.36 diopters. We also measured discrimination thresholds and confirm that the just noticeable difference between two objects at different relative depths in front of a background stimulus is directly proportional to their absolute depth from the background. These thresholds confirm that ocular parallax is an import \emph{visual cue} in VR. Our third and fourth experiments show that ocular parallax acts as reliable ordinal depth cue but that it may not be a reliable absolute depth cue. Finally, our fifth experiment demonstrates that ocular parallax rendering significantly improves the impression of realistic depth when viewing a 3D scene over conventional rendering. 

\paragraph{\bf Ocular Parallax and Retinal Blur} Retinal blur describes the depth-dependent defocus blur of objects on the retina, relative to the accommodation distance. In physical environments, retinal blur discrimination thresholds are comparable to those of ocular parallax---approximately $\pm$0.35~D~\cite{Ogle:59}---and have been shown to serve as both a reliable ordinal and absolute depth cue~\cite{Vishwanath:10}. While recent research efforts in VR/AR have studied focus cues such as retinal blur, chromatic aberrations, and accommodation, none have studied ocular parallax, even though their effect sizes are very similar.


Gaze-contingent retinal blur and ocular parallax rendering can certainly be implemented simultaneously as they are complementary cues. While depth-of-field rendering blurs objects at different depths, we do not expect it to alter our measurements or insights significantly. This is because the perception of motion is understood to use low spatial frequencies~\cite{Smith:94}, so the loss of the high spatial frequencies due to depth-of-field rendering should not impair motion, or ocular parallax, perception. Indeed, our measured detection thresholds are consistent with those measured in a physical environment~\cite{Bingham:1993}, where retinal blur was apparent.  

\paragraph{\bf Eye Model} Throughout the paper, we use a schematic eye that makes several simplifying assumptions. First, we assume that the front nodal point is indeed the center of projection. Second, we assume that the user does not accommodate. Third, we assume that the optical and visual axis of the eye are the same. The first assumption is reasonable and in line with the Gullstrand-Emsley schematic eye. However, it could be argued that there is no exact center of projection in the eye, that it varies significantly between users, or that it is the center of the entrance pupil of the eye, which laterally shifts with changing pupil diameter~\cite{atchison2014effects}, instead of the front nodal point. The precise location of the center of projection is a topic that deserves further discussion and that should also be experimentally located, which we leave for future work. The second assumption requires the user to accommodate at a fixed, far distance. For near-eye displays that support accommodation, vergence or accommodation tracking could be used to model the accommodation-dependent nodal point. Finally, we assume that optical and visual axis of the eye are the same. Using this assumption, we demonstrate that ocular parallax rendering has no significant effect on disparity distortion, thus absolute depth perception, in Section~\ref{sec:depth}. However, further studies on this topic should be conducted using more accurate eye models that include, for example, an offset between visual and optical axis of the eye.

\paragraph{\bf Limitations} Although our system uses some of the highest-end components available, including a high-resolution wide-field-of-view VR display and a 120~Hz eye tracker, the latency of 20~ms for gaze-contingent ocular parallax rendering is high. Faster and more accurate eye tracking would certainly help improve the user experience for all gaze-contingent rendering schemes, including ocular parallax. 

\paragraph{\bf Applications} We envision ocular parallax rendering to be a standard part of the graphics pipeline of eye-tracking-enabled near-eye displays. It improves perceptual realism, ordinal depth perception, and it may offer other perceptual benefits. In particular, optical see-through augmented reality systems may benefit from ocular parallax rendering as a user usually sees a digitally rendered stimulus overlaid on a reference stimulus (i.e. the physical world); visual cue consistency between these stimuli may be even more important than in VR. Finally, one could also imagine that an amplified version of ocular parallax rendering could be an effective gaze-contingent user interface that allows users to transform objects, navigate through virtual environments, or perform other tasks. Tasks that require hands-free operation could particularly benefit from this type of gaze-contingent interaction mode. 


\paragraph{\bf Future Work} As the fields of view of emerging VR and AR systems keep increasing, understanding perceptual effects in peripheral vision becomes ever more important. With this work, we thoroughly evaluate the perceptual implications of one technique, ocular parallax rendering, which shows the strongest effects in near-mid peripheral vision. However, many other technologies, such as foveated rendering, reducing motion sickness, or other means to improve depth perception or perceptual realism, could also benefit from studying their perceptual effects on peripheral vision. Additionally, as many of these effects vary from person to person, performing a per-user calibration of the distance between the centers of rotation and projection could further increase perceptual realism.

\paragraph{\bf Conclusions} Virtual and augmented reality systems have focused on improving resolution, field of view, device form factor, and other characteristics. With this work, we hope to stimulate new directions for gaze-contingent rendering and improve perceptual realism and depth perception with next-generation near-eye displays. 

\section*{Acknowledgments}
\label{sec:acknowledgments}
This project was generously supported by funding from the National Science Foundation (NSF, award numbers 1553333 and 1839974), a Sloan Fellowship, an Okawa Research Grant, and Intel.

\bibliographystyle{ACM-Reference-Format}
\bibliography{ms}

\end{document}


\title{Supplementary Information:\\ Gaze-Contingent Ocular Parallax Rendering for Virtual Reality}

\author{Robert Konrad}
\affiliation{%
  \institution{Stanford University}
}
\email{rkkonrad@stanford.edu}
\author{Anastasios Angelopoulos}
\affiliation{%
  \institution{Stanford University}
}
\email{nikolasa@stanford.edu}
\author{Gordon Wetzstein}
\affiliation{%
  \institution{Stanford University}
}
\email{gordon.wetzstein@stanford.edu}




\maketitle

In this document we provide additional discussion and results in support of the primary text.

\section{Ocular Parallax Linearity in Dioptric Space}

\begin{figure}[h]
	\centering
		\includegraphics[width=0.8\textwidth]{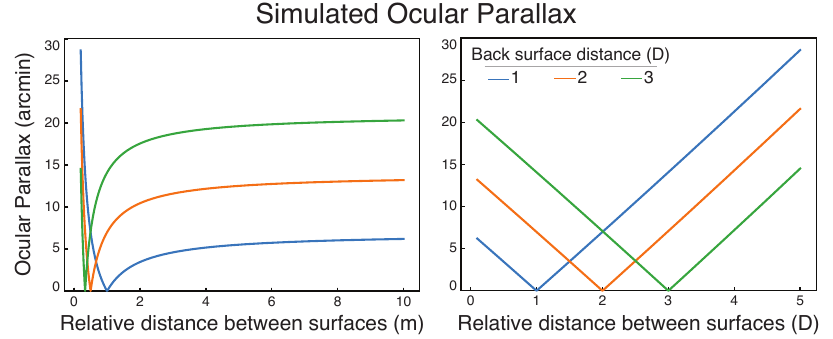}
		\caption{Model of ocular parallax in metric and dioptric spaces. The plots show the estimated amount of ocular parallax between two objects separated in depth. The differently colored lines indicate the ocular parallax when the back object is set to 1, 2, or 3~D for varying distances to the front object. The left plot shows that the ocular parallax is not linear with metric distance, is linear with dioptric distance. }
		\label{fig:supp:linearity}
\end{figure}

The amount of parallax due to eye rotations that is observed near an occlusion boundary is a function of the magnitude of the eye rotation, the distance to the farther surface, and the separation between two surfaces. \citet{Bingham:1993} presents a model taking these factors into account and the ocular parallax observed when the distances are considered in metric units is shown in Figure~\ref{fig:supp:linearity} (left). Clearly, ocular parallax is not linear with separation distance in meters. However, the amount of ocular parallax observed on the retina is linear when the distances are considered in diopters, or inverse meters (Figure~\ref{fig:supp:linearity}, right). The simulated ocular parallax results in Figure~\ref{fig:supp:linearity} assumed 15\textdegree\ of eye rotation and a distance between the nodal point and center of rotation of 7.69~mm.

\newpage
\section{Eye Tracker Characterization}

\begin{figure} [h]
	\centering
		\includegraphics[width=\textwidth]{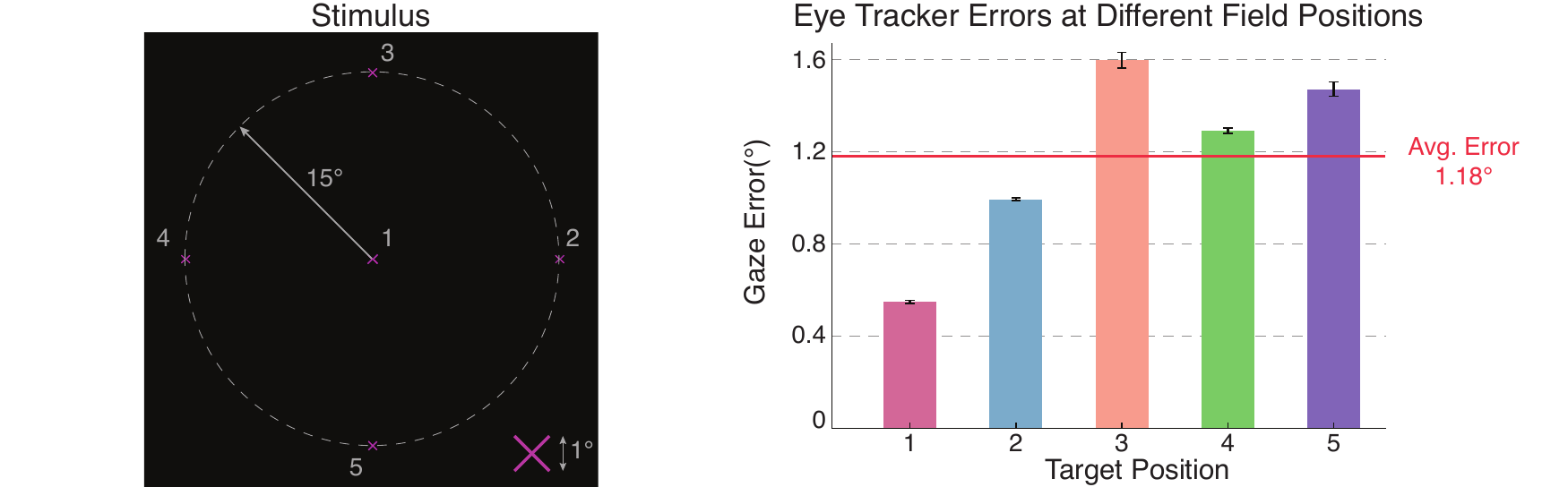}
		\caption{Eye tracker characterization. After calibrating the eye tracker, subjects viewed each of the 5 targets on the left and their gaze data was recorded. The average errors across subjects for each field position is shown on the right. Naturally, the lowest error was observed at the center of the visual field, but the error never rose above 1.6\textdegree\ (target position \#3) at the measured field positions.}
		\label{fig:eye_tracker}
\end{figure}

We characterized the Pupil Labs eye tracker accuracy on a subset of the subjects from the experiments in the main paper (4 subjects, ages 22--28, 1 female). Subjects performed the manufacturer-provided 7-point eye tracker calibration followed by an additional single-point eye-tracker re-calibration to mimic the procedure of the main experiments. Subjects then viewed each of the five \scalebox{1.25}{$\times$}$\!$-shaped targets in the order shown in Figure~\ref{fig:eye_tracker} (left); each target subtended 1\textdegree\ of visual field. The targets were placed at the center of their visual field and 15\textdegree\ above, below, to the left, and to the right of the center. The subjects were instructed to look to the center of the targets and indicated they were doing so with a keyboard press which recorded 5 seconds of gaze data. Users repeated this procedure for each target. The accuracy computation assumed that the user looked at the center of the target.

The results of the eye tracker accuracy characterization can be found in Figure~\ref{fig:eye_tracker} (right). The error averaged across all subjects and field positions was 1.18\textdegree, which is worse than the 1.0\textdegree\ error claimed by Pupil Labs. Clearly, the eye tracker error increases with increasing eccentricity, but was never higher than 1.6\textdegree\ (for the field position \#3). One subject wore contacts which is known to deteriorate eye tracker performance. Indeed, the eye-tracker accuracy for this subject was 1.415\textdegree\ while the average eye-tracker accuracy for the subjects without contacts was 1.1\textdegree.

\newpage
\section{Psychometric Functions}

\begin{figure}[h]
	\centering
		\includegraphics[width=0.91\textwidth]{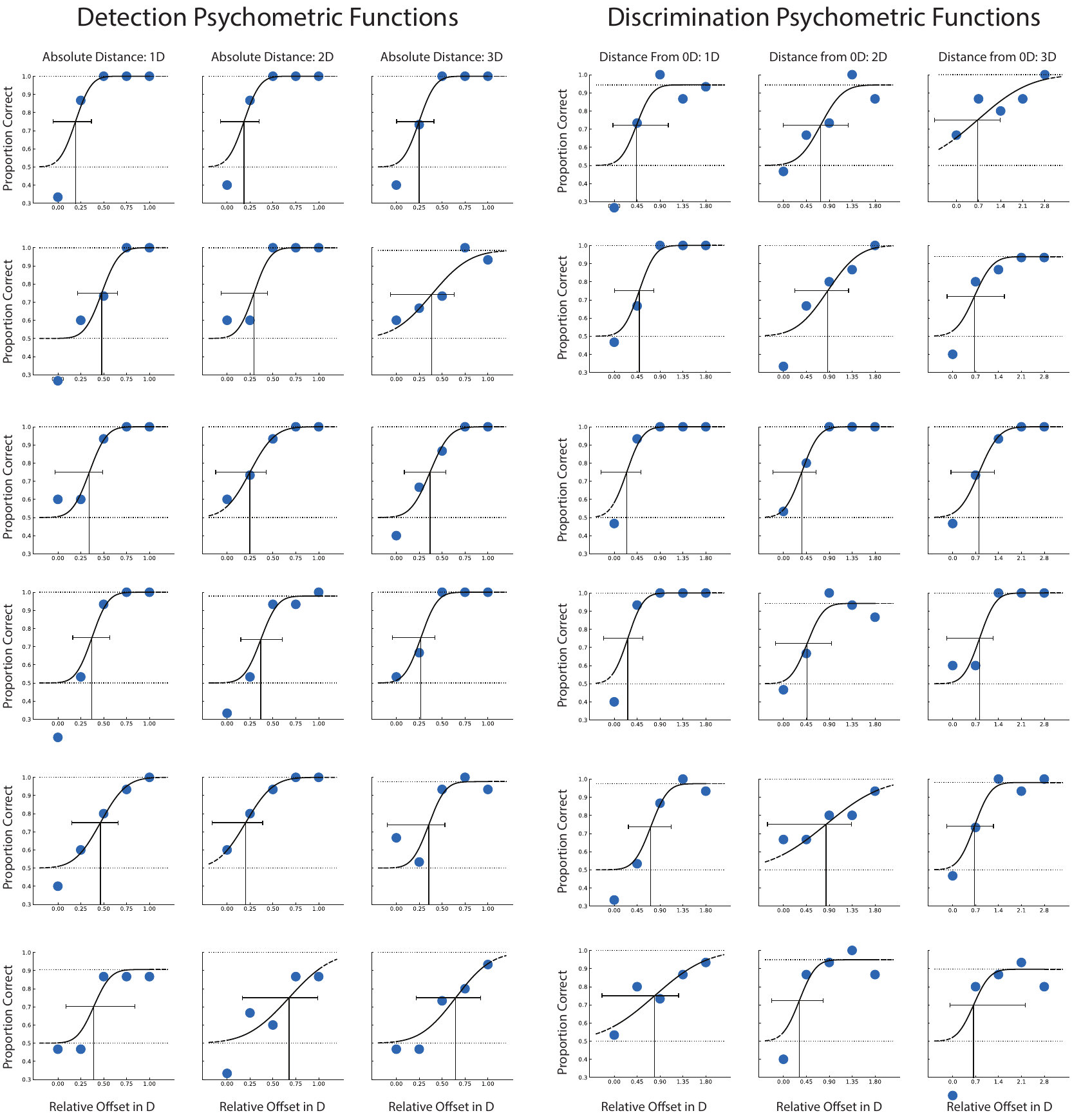}
		\vspace{-15pt}
		\caption{Detection and Discrimination psychometric functions. The left column shows the psychometric functions for the ocular parallax detection experiment, while the right column presents the functions for the ocular parallax discrimination experiment. Each row of each column corresponds to the psychometric functions measured for one subject for 3 different stimulus conditions. The vertical lines intersecting the psychometric fit indicate the threshold corresponding to the 75\% correct response rate, and the horizontal line abutting it representing the 95\% confidence interval. Lower thresholds correspond to subjects being more sensitive to the ocular parallax effect.}
		\label{fig:supp:psychometrics}
\end{figure}

We present the full set of psychometric functions from our detection and discrimination threshold experiments in \autoref{fig:supp:psychometrics}. The \textit{psignifit} Python package~[Sch\"utt et al. 2016; Wichmann and Hill 2001a,b], was used for fitting psychometric functions to the data using Bayesian inference. 
\vspace{-7pt}

\newpage
\section{Additional Details on Egocentric Depth Study}

Multiple reports have shown that egocentric depth perception, or the subjective perceived distance to objects around one's self, is shorter in virtual environments than in real ones~\cite{Renner:2013}. Factors like vergence and accommodation~\cite{Watt:2005}, the quality of graphics~\cite{Knapp:03}, and even the weight and inertia of the HMD itself have been deemed to contribute to this effect. Ocular parallax rendering could reduce the underestimation by minimizing disparity distortions and providing additional depth information through micro parallax and gaze-contingent occlusions. We investigate participants' egocentric depth estimation in their action or personal space (i.e. within arm's reach), where the ocular parallax effect is strongest. Because verbal estimates have been shown to be variable~\cite{Napieralski:11}, we rely on a blind reaching task where users view a target and then, with the display turned off, reach to where it was last seen. The blind reaching task is the standard way to evaluate egocentric distance estimation for objects within arm's reach~\cite{Altenhoff:12}.

Sixteen volunteers participated in the study (age range 22-32, 5 females), of which one was excluded for not passing a standard Randot stereo vision test, and two others were excluded due to the eye tracker failing to track their pupils.  

\begin{figure}[b]
	\centering
		\includegraphics[width=0.75\columnwidth]{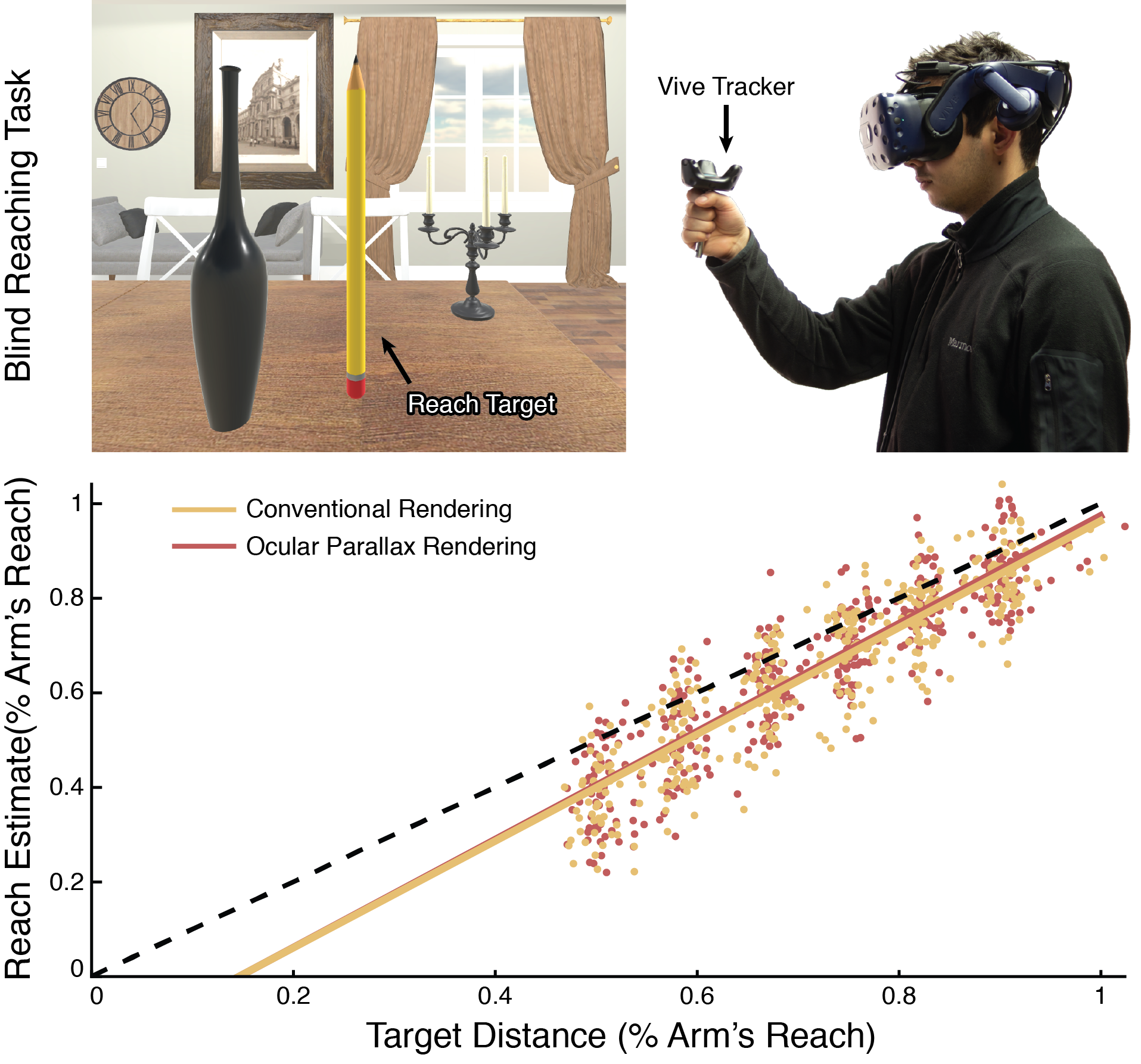}
		\caption{Egocentric depth perception. This study investigates whether the additional depth cues in ocular parallax aid egocentric depth perception. The viewing environment including the reach target, a pencil, are shown in the top left. Users viewed the target and then, with the display turned off, reached to it as seen in the top right. Participants' reaches in the ocular parallax enabled and disabled modes are found in the bottom plot.}
		\label{fig:egocentric}
\end{figure}

\paragraph{\bf Stimuli and study setup} To minimize any potential effects of graphics quality on depth judgment~\cite{Knapp:03}, the viewing environment for the blind reaching task consisted of a photo-realistic dining room scene as seen in \autoref{fig:egocentric}. A floating pencil served as the reaching target and was surrounded by objects at different depths and a feature rich background, emphasizing the ocular parallax effect. 

The HTC Vive Pro with Pupil Labs eye trackers served as the HMD; 6-DOF head pose tracking was enabled. The hand's position was tracked via an HTC Vive tracker mounted on an optical post that was held upright (\autoref{fig:egocentric}, top right). To facilitate natural reaching, participants held the tracker in their dominant hand and an HTC Controller in their other hand for interaction.

\paragraph{\bf Conditions} Each participant performed the blind reaching task with ocular parallax enabled and disabled. To maintain a common reach metric between participants, reach target distances were set proportionally to each participants' arm reach. The reach targets appeared at 50, 58, 67, 75, 82, and 90\% of the participant's maximum arm reach. The participants completed trials under each of the 12 conditions 5 times and the presentation order of the 60 trials was randomized.

\paragraph{\bf Procedure} Each participant performed an ipd, eye tracker, and maximum arm reach calibration procedure at the beginning of the session. To determine their maximum arm reach, participants were instructed to fully extend their arm directly in front of their head. The distance between the eyes' midpoint and the hand defines the maximum arm reach. 

A standard blind reaching task was then conducted~\cite{Altenhoff:12}. Participants first observed the viewing environment to familiarize themselves with its size and object depths. They then began the trials. Each participant performed two practice trials followed by 60 recorded distance estimates with each trial consisting of a target presentation phase and a blind reaching phase. In the final presentation phase, the pencil appeared in front of the users and they were instructed to determine its depth by looking at it, but also to gaze to other scene objects to maximize eye rotation and the ocular parallax effect. Participants were required to view the environment for a minimum of 5 seconds before they were given the ability to turn off the display when they were ready to perform the blind reach. With the display off, they were instructed to reach for the pencil with only their arm while their torso remained steady. They progressed onto the next trial by pulling the controller's trigger when satisfied with their reach position. All head and hand pose tracking data, as well as the pencil's position, were recorded during the blind reaching phase.

\paragraph{\bf Analysis} The reach estimate was computed as the average of the last 10 samples of the participant's reach data. Because 6-DOF head tracking was enabled, natural head movement resulted in deviation from the the head to target distances reported above. To account for this, the data was analyzed along the horizontal 2D vector between the eyes' midpoint and the target's position. Both the target's distance and reach distance were computed along this vector as a percentage of the arm's maximum reach. 

\paragraph{\bf Results} Enabling ocular parallax rendering had little to no effect on the egocentric depth perception of the participants. \autoref{fig:egocentric} shows each participant's reach, as a percentage of their maximum arm reach, to a presented target distance for the ocular parallax enabled and disabled conditions. Linear models were fit to the two sets of reaches, finding that the slopes for the ocular parallax enabled and disabled modes were 1.141 and 1.128, respectively, while the intercepts were 0.167 and 0.165, respectively. A multiple regression analysis did not show a significant difference between the two conditions. 

\bibliographystyle{ACM-Reference-Format}
\bibliography{supplement}